\documentclass[preprint,12pt]{elsarticle}
\usepackage{amsmath,amssymb}
\usepackage{graphicx}
\usepackage{booktabs}
\usepackage{multirow}
\usepackage{array}
\usepackage{hyperref}
\usepackage{lineno}
\usepackage{adjustbox}
\usepackage{placeins}
\biboptions{sort&compress}

\journal{Physics of the Dark Universe}

\begin{document}

\begin{frontmatter}

\title{Interacting dark energy constraints from \textit{Fermi} GRBs and Pantheon+ SNe Ia with full GRB covariance}

\author[aff1]{Jianfeng Meng}
\author[aff2]{Xiaofeng Yang\corref{cor1}}
\ead{xfyang@henu.edu.cn}
\author[aff2]{Yunliang Ren}
\author[aff2]{Yangjun Shi}
\author[aff2]{Bohao Wang}
\author[aff2]{Jingze Li}
\author[aff1]{Xiongwei Liu}
\cortext[cor1]{Corresponding author}

\affiliation[aff1]{organization={School of Physics and Astronomy, China West Normal University},
            city={Nanchong},
            postcode={637002}, 
            country={China}}
\affiliation[aff2]{organization={School of Physics and Electronics, Henan University},
            city={Kaifeng},
            postcode={450046}, 
            country={China}}

\begin{abstract}
The standard $\Lambda$CDM model faces long-standing theoretical and observational problems, such as the Hubble tension, which motivate extensions beyond $\Lambda$CDM, including interacting dark energy (IDE). Type Ia supernovae (SNe Ia) are precise probes of the late-time expansion history, while gamma-ray bursts (GRBs) can extend the Hubble diagram to higher redshifts. However, GRB cosmology depends on careful calibration and uncertainty modeling. Using an Amati relation calibrated with the corresponding low-redshift GRBs, we construct distance moduli for the high-redshift subsets of the 15-year \textit{Fermi}/GBM GOLD and FULL samples and combine them with Pantheon+ SNe Ia to compare flat $\Lambda$CDM, $w$CDM, IDE-$\rho_{\rm de}$, and IDE-$\rho_{\rm c}$ models. The covariance of the calibrated Amati intercept and slope is propagated into a full, non-diagonal GRB distance-modulus covariance, and an effective residual scatter, $\sigma_{\rm res,\mu}$, is fitted jointly with the cosmological parameters. The GOLD and FULL samples yield very similar constraints on the main cosmological parameters. With either the full or diagonal GRB covariance, the IDE models do not improve likelihood sufficiently to compensate for their additional parameters, and the BIC favors $\Lambda$CDM. A SNe-only comparison shows that the additional constraining power of the GRBs is generally modest. Current GRB and Pantheon+ distance measurements provide no significant evidence for either of the two IDE interactions considered here.
\end{abstract}

\begin{highlights}
\item A full GRB covariance matrix propagates the correlated Amati-calibration uncertainty.
\item An effective residual scatter in GRB distance-modulus space is sampled as a nuisance parameter.
\item The IDE conclusion is stable under full and diagonal GRB covariances.
\item Current GRB and Pantheon+ data provide no evidence for interacting dark energy.

\end{highlights}


\end{frontmatter}

\section{Introduction}
In the late 1990s, observations of Type Ia supernovae (SNe Ia) provided the first direct evidence that the expansion of the universe is accelerating~\cite{Riess,Perlmutter}. This fact was later confirmed by many independent observations~\cite{Weinberg}, such as anisotropies in the cosmic microwave background (CMB)~\cite{Spergel,Komatsu,Planckk,Planck2018,Madhavacheril}, baryon acoustic oscillations (BAO)~\cite{Eisenstein,Percival,Beutler,Parkinson,Alam1,Alam2}, inhomogeneities in the matter distribution~\cite{Cole,Tegmark}, and weak gravitational lensing~\cite{Bartelmann,Contaldi,Liu,Shan,Mandelbaum,Heymans,Abbott}. The accelerated expansion of the universe is attributed to a mysterious component called dark energy. In the standard $\Lambda$CDM cosmological model, dark energy is described simply as a component with constant energy density, with the equation of state $w_{\Lambda} \equiv \frac{p_{\Lambda}}{\rho_{\Lambda}} = -1$~\cite{Peebles,Yoo}.

Although the $\Lambda$CDM model can explain many astronomical and cosmological observations, the physical nature of its two largest components, dark energy and dark matter, is still unclear. The model also faces several long-standing theoretical issues, such as the cosmological constant problem~\cite{Weinbergg} and the coincidence problem~\cite{Chimento}. Observationally, $\Lambda$CDM is also challenged by several tensions, most notably the Hubble tension~\cite{Valentino1}, which refers to the discrepancy between the Hubble constant inferred from early-universe observations under the assumption of $\Lambda$CDM and that measured directly from late-universe probes~\cite{Planck2018,Riesss}. Recent local-distance-network measurements suggest that this discrepancy has reached the $\sim5\sigma$ level, and can be even larger for some CMB-based early-universe references~\cite{H0DN}, making it one of the most severe challenges to the standard cosmological model. These problems suggest that new physics beyond the current standard cosmological model may be needed. For this reason, $w$CDM and interacting dark energy (IDE) models have been proposed. A simple extension of $\Lambda$CDM is $w$CDM, in which the equation of state of dark energy can be a constant different from $-1$. Another class of extensions is provided by IDE models, which allow energy exchange between dark energy and dark matter. These models have their theoretical roots in early studies of non-minimally coupled scalar fields, coupled quintessence, and interacting scalar-field cosmology~\cite{Amendola1,Amendola2,Billyard}. Such an interaction may change the late-time expansion history and help alleviate the coincidence problem and the Hubble tension~\cite{Amendola2,Zimdahl,Costaa,Pann,Pan,Di,Abdallaa}. IDE can be studied phenomenologically with only minimal assumptions about the underlying theory~\cite{2011MNRAS.416.1099C}. For example, one may assume that the ratio between the dark energy and matter densities follows a power law of the scale factor~\cite{2001PhRvL..87n1302D}, $r \equiv \rho_{\rm de}/\rho_{\rm m} \propto a^{\xi}$, where $\rho_{\rm de}$ and $\rho_{\rm m}$ are the energy densities of dark energy and matter, respectively, and the free parameter $\xi$ measures the interaction strength and can be determined from observations. We can also assume that the interaction term is a linear combination of the dark energy and dark matter densities, $Q = 3H(\xi_1 \rho_{\rm de} + \xi_2 \rho_{\rm c})$, which includes the commonly studied cases proportional to the dark energy density, $Q = 3H\xi\rho_{\rm de}$, to the dark matter density, $Q = 3H\xi\rho_{\rm c}$, or to the total dark-sector density, $Q = 3H\xi(\rho_{\rm de}+\rho_{\rm c})$~\cite{Amendola2,Billyard,Zimdahl,Chimento,viita,Clemson}. These interaction models have been widely studied~\cite{Pan,PanSupriya,Valentino,Yang,Escamilla,Bensity,Giare,Litiannuo,Nong,Silva,Hoerning,Zhu,Figueruelo}. However, these models also introduce additional parameters. The key question is whether IDE models improve the fit sufficiently to offset the penalty associated with their additional parameters.

SNe Ia are among the most important probes of the late-time expansion history of the universe. They provide a precise Hubble diagram and cover a wide redshift range. However, the redshift coverage of current SNe Ia samples is limited at high redshift. Gamma-ray bursts (GRBs) can be observed at much higher redshifts and can extend late-time cosmological tests into a high-redshift regime that is only sparsely sampled by SNe Ia. Nevertheless, GRBs are not standard candles and must be standardized through empirical luminosity relations before they can be used for cosmology~\cite{Amati2001,Ghirlanda,Yonetoku,Schaefer,Dainotti2008}. One commonly used empirical relation is the Amati relation~\cite{Amati2001}, which describes the relation between the spectral peak energy in the GRB cosmological rest frame, $E_{\rm p}$, and the isotropic energy, $E_{\rm iso}$. The reliability of GRBs as distance indicators depends on the calibration of the empirical relation, the construction of the covariance matrix, and the treatment of intrinsic scatter.

In addition to statistical measurement errors, several astrophysical and observational systematics may affect GRB cosmology~\cite{Butler,Ghirlan,LiLi,Schady,Perley}. These include selection effects caused by detector thresholds and sample cuts, possible redshift evolution of the luminosity relation, uncertainties in detector response, spectral fitting, and bolometric-fluence estimates, incompleteness in redshift measurements, host-galaxy extinction, and possible dependence on the burst environment or progenitor metallicity. The possible impact of selection effects on GRB spectral-energy correlations has been discussed in detail in the literature, although its importance is still debated. Possible redshift evolution of the Amati relation has also been tested in several studies, and host-galaxy dust or population effects can affect the observed GRB sample. In this work, we control the calibration uncertainty by propagating the covariance of the Amati-relation parameters and by sampling an additional effective residual scatter. We do not attempt a full population-level model of all GRB systematics. Such a treatment is beyond the scope of this paper and is left for future work.

Based on 15 years of the \textit{Fermi}/GBM (Gamma-ray Burst Monitor) catalog, Wang \& Liang~\cite{2024MNRAS.533..743W} constructed updated GOLD and FULL samples of long GRBs spanning $0.0785 \leq z \leq 8.2$. The GOLD sample contains 123 long GRBs with $z \leq 5.6$, and the FULL sample contains 151 long GRBs with $z \leq 8.2$. They used low-redshift GRBs and Gaussian-process reconstructions of 32 cosmic-chronometer measurements to calibrate the Amati relation with D'Agostini, Reichart, and Bivariate Correlated Errors and intrinsic Scatter (BCES) methods, and then combined the high-redshift subsets with Pantheon+ SNe Ia for cosmological constraints. Cao $\&$ Ratra~\cite{CaoRatra}, using a simultaneous correlation-cosmology analysis of the same 151-GRB compilation, found that the matter-density constraints from the A123 and A123+A28 samples are in more than $2\sigma$ tension with better-established $H(z)$+BAO constraints in most of their models. Their result highlights the need to assess the robustness of the cosmological constraints derived from this GRB compilation and motivates explicit tests of the GRB error model and the additional constraining power provided by GRBs. Recent related studies include an artificial-neural-network test of the cosmic distance duality relation with GOLD/FULL GRBs, Pantheon+, DESI DR2, and strong lenses~\cite{Xie}; a phenomenological interacting-dark-energy analysis combining the high-redshift \textit{Fermi} samples with DESI DR2 BAO~\cite{Zhu}; and a redshift-binned reconstruction based on the Combo correlation calibrated from 244 GRB data with updated values of spectral peak energy including the \textit{Fermi} catalog~\cite{2024MNRAS.533..743W} that found no significant evidence for dynamical dark energy~\cite{Marco}. These studies demonstrate the potential of GRBs as high-redshift cosmological probes, while showing that the resulting constraints can depend on the calibration method and data combination.

In this work, we propagate the full covariance of the BCES-Orthogonal Amati-relation intercept and slope~\cite{2024MNRAS.533..743W} into the high-redshift GRB distance-modulus covariance matrix. An effective residual scatter in distance-modulus space, $\sigma_{\rm res,\mu}$, is introduced and sampled jointly with the cosmological parameters. We test whether current GRB and Pantheon+ data favor IDE models over $\Lambda$CDM. We further examine whether the model ranking is robust to the choice between the full and diagonal GRB covariance treatments and how this choice affects the inferred value of $\sigma_{\rm res,\mu}$.

\section{Phenomenological interacting dark energy models}
We describe the homogeneous and isotropic universe with a flat FLRW metric:
\begin{equation}
ds^2 = -dt^2 + a^2(t)\left[dr^2 + r^2 d\theta^2 + r^2 \sin^2\theta \, d\phi^2\right].
\end{equation}
The continuity equation for each component in the universe is:
\begin{equation}
\dot{\rho} + 3\frac{\dot{a}}{a}(1+w)\rho = 0.
\end{equation}
where $a(t)$ is the scale factor at cosmic time $t$, with its present value set to $a(t_0)=1$, 
$\rho = \sum \rho_i$ is the total energy density, and $w = p/\rho$ is the equation-of-state parameter.

In standard cosmology, dark energy and dark matter are assumed to evolve independently and are separately conserved. In IDE models, the two dark sectors are coupled through a nongravitational energy–momentum transfer. Their individual energy densities are no longer conserved separately, although the total energy density of the coupled dark sector remains conserved. The continuity equations are written as:
\begin{align}
\dot{\rho}_{\rm b} + 3H\rho_{\rm b} &= 0, \\
\dot{\rho}_{\rm c} + 3H\rho_{\rm c} &= Q, \\
\dot{\rho}_{\rm de} + 3H(1+w)\rho_{\rm de} &= -Q .
\end{align}
The subscripts b, c, and de denote baryons, cold dark matter, and dark energy, respectively. The interaction term $Q$ describes the energy transfer between the dark sectors. With the sign convention adopted here, $Q > 0$ corresponds to energy transfer from dark energy to dark matter, while $Q < 0$ corresponds to energy transfer from dark matter to dark energy.

In this work, we consider two widely studied interaction forms: one proportional to the dark energy density, $Q = 3H\xi\rho_{\rm de}$, and one proportional to the dark matter density, $Q = 3H\xi\rho_{\rm c}$. We refer to them as IDE-$\rho_{\rm de}$ and IDE-$\rho_{\rm c}$, respectively. The dimensionless parameter $\xi$ quantifies the interaction strength. When $\xi = 0$, the model reduces to the non-interacting case.

For a flat late-time universe, the Friedmann equation is,
\begin{equation}
H^2(z) = \frac{8\pi G}{3}\left[\rho_{\rm m}(z) + \rho_{\rm de}(z)\right],
\end{equation}
The background expansions for IDE-$\rho_{\rm de}$ and IDE-$\rho_{\rm c}$ are, respectively,
\begin{align}
E^2(z) &= 
\left[\Omega_{\rm m0} + \frac{\xi}{w+\xi}(1-\Omega_{\rm m0})\right](1+z)^3
+ \frac{w}{w+\xi}(1-\Omega_{\rm m0})(1+z)^{3(1+w+\xi)}, \\
E^2(z) &= 
\Omega_{\rm b0}(1+z)^3
+ \Omega_{\rm c0}\frac{w}{w+\xi}(1+z)^{3(1-\xi)}
+ \left(1-\Omega_{\rm b0}-\frac{w}{w+\xi}\Omega_{\rm c0}\right)(1+z)^{3(1+w)} .
\end{align}
where $\Omega_{\rm m0} = \Omega_{\rm b0} + \Omega_{\rm c0}$ and 
$\Omega_{\rm de0} = 1 - \Omega_{\rm m0}$.

\section{Data and methodology}
\subsection{Gamma-ray burst data}
We use the updated \textit{Fermi}/GBM long-GRB sample constructed by Wang \& Liang~\cite{2024MNRAS.533..743W}, including the GOLD and FULL samples. The GOLD sample contains 123 GRBs with $z \leq 5.6$, and the FULL sample contains 151 GRBs with $z \leq 8.2$.

The Amati relation~\cite{Amati2001} for GRBs describes the relation between the spectral peak energy in the GRB cosmological rest frame, $E_{\rm p}$, and the isotropic energy, $E_{\rm iso}$,
\begin{equation}
\log_{10}\left(\frac{E_{\rm iso}}{1~\mathrm{erg}}\right)
= a + b \log_{10}\left(\frac{E_{\rm p}}{300~\mathrm{keV}}\right),
\end{equation}
where $a$ and $b$ are parameters to be determined from observations, and
\begin{equation}
E_{\rm p} = E_{\rm p} ^ {obs}(1+z), \qquad
E_{\rm iso} = 4\pi d_L^2 S_{\rm bolo}(1+z)^{-1}.
\end{equation}

Wang \& Liang calibrated the $z<1.4$ subsets using a Gaussian-process reconstruction of 32 observational $H(z)$ measurements obtained with the cosmic-chronometer method~\cite{2024MNRAS.533..743W}. Their comparison included D'Agostini, Reichart, BCES($Y|X$), and BCES-Orthogonal fits. We adopt their BCES-Orthogonal calibration results because neither $E_{p}$ nor $E_{\rm iso}$ can unambiguously be treated as the independent variable~\cite{Akritas,Hogg}. The adopted values are $(a,b)=(52.43,2.16)$ for GOLD and $(52.39,2.46)$ for FULL, together with their published $2\times2$ covariance matrices.

\subsection{GRB distance modulus, covariance, and effective residual scatter}
For each GRB, the required observables include the redshift $z$, the rest-frame spectral peak energy $E_{\rm p,i}$ and its uncertainty, the measured bolometric fluence $S_{\rm bolo}$ and its uncertainty.

The luminosity distance and distance modulus of GRBs are, respectively,
\begin{equation}
    d_{L,\rm GRB} = \left[\frac{(1+z)E_{\rm iso}}{4\pi S_{\rm bolo}}\right]^{1/2},
\end{equation}

\begin{equation}
    \mu_{\rm GRB,obs}
=5\log_{10}\left(\frac{d_{L,\rm GRB}}{\rm Mpc}\right)+25.
\end{equation}

The baseline propagated covariance is
\begin{equation}
C_{\rm GRB} = C_{\rm meas} + C_{\rm cal},
\end{equation}
where $C_{\rm meas}$ is the measurement error term, obtained by propagating the errors in $E_{\rm p,i}$ and $S_{\rm bolo}$. The second term, $C_{\rm cal}$, is the calibration covariance propagated from the $(a,b)$ covariance. Because the same calibrated $a$ and $b$ affect every high-redshift GRB, $C_{\rm cal}$ is non-diagonal.

The covariance of the Amati calibration parameters $(a,b)$ describes the uncertainties in the calibrated intercept and slope, but it does not fully describe the additional burst-to-burst dispersion. We add an extra scatter
\begin{equation}
C_{\rm eff} = C_{\rm GRB} + \sigma_{\rm res,\mu}^{2} I .
\end{equation}

The fitted $\sigma_{\rm res,\mu}$ is an effective residual scatter in GRB distance modulus space. It absorbs any remaining burst-to-burst dispersion not represented by the measurement and calibration covariance. This may include genuine Amati dispersion as well as selection effects, redshift evolution, unmodeled spectral or fluence errors, lensing scatter, and other systematics. However, it is not intended to provide a detailed physical model of all possible GRB systematics.

The parameter $\sigma_{\rm res,\mu}$ is treated as a nuisance parameter and sampled jointly with the cosmological parameters. This follows the standard treatment of empirical astronomical correlations with extra variance, where the additional scatter is included as a model parameter and marginalized over~\cite{Agostini,Hogg}. It enters the GRB likelihood through the effective covariance matrix $C_{\rm eff}$~\cite{Heavens2009}. The full-covariance likelihood is written as:
\begin{equation}
-2\ln \mathcal{L}_{\rm GRB}
= \Delta\mu_{\rm GRB}^{T} C_{\rm eff}^{-1} \Delta\mu_{\rm GRB}
+ \ln |C_{\rm eff}| .
\end{equation}

The cosmological likelihood uses the high-redshift subsets: 60 GOLD GRBs over $1.405\leq z\leq5.6$ and 78 FULL GRBs over $1.4\leq z\leq8.2$. For the diagonal diagnostic we retain all diagonal variances but discard the correlations,
\begin{equation}
C_{\rm eff}^{\rm diag}
=
{\rm diag}(C_{\rm GRB})
+
\sigma_{\rm res,\mu}^{2} I,
\end{equation}

This comparison is motivated by the fact that many GRB cosmological analyses adopt diagonal errors and allows us to assess whether the calibration-induced correlations affect the cosmological conclusions.

\subsection{Pantheon+ SNe Ia data}
We perform a joint analysis of the GRB and Pantheon+ SNe Ia data sets ~\cite{Scolnic,Brout}to constrain the cosmological models. The Pantheon+ sample comprises 1701 SNe Ia light curves from 1550 distinct SNe Ia. The SNe Ia contribution to the total $\chi^2$ is
\begin{equation}
\chi^2_{\rm SNe}
= \Delta\mu_{\rm SNe}^{T} C_{\rm SNe}^{-1} \Delta\mu_{\rm SNe},
\end{equation}
where $\Delta\mu_{\rm SNe} = \mu_{\rm SNe,obs} - \mu_{\rm th}$ is the difference between the observed and theoretical distance moduli, and $C_{\rm SNe}$ is the full covariance matrix including statistical and systematic uncertainties~\cite{Brout}.

\subsection{Cosmological inference and model comparison}
We compare four cosmological models: $\Lambda$CDM, $w$CDM and two interacting models, IDE-$\rho_{\rm de}$ and IDE-$\rho_{\rm c}$. We require the background evolution to remain physical over the range considered. The specific requirements are:
\begin{align}
E^2(z) &> 0, \\
\rho_{\rm c}(z) &> 0, \\
\rho_{\rm de}(z) &> 0 .
\end{align}

We sample $H_0$, $\Omega_{\rm c0}$, $\omega_{\rm b}$, and $\sigma_{\rm res,\mu}$. Here $\omega_{\rm b} \equiv \Omega_{\rm b}h^2$, where $h = H_0/100$. The baryon density parameter used in the background evolution is computed from $\Omega_{\rm b0} = \omega_{\rm b}/h^2$, and the present total matter density is given by $\Omega_{\rm m0} = \Omega_{\rm b0} + \Omega_{\rm c0}$.

Since the present analysis is restricted to late-time background probes, which do not constrain the baryon density competitively on their own, we impose a Gaussian prior on the physical baryon density from the Planck 2018 results, adopting  $\omega_{\rm b} = 0.02237 \pm 0.00015$ as an external reference motivated by precision CMB measurements~\cite{Planck2018}. This choice is common in cosmological fitting, where the baryon density is often tightly constrained or fixed using CMB information to break degeneracies in the matter sector~\cite{Alfano2024}. For $w$CDM and IDE models, we also sample the constant dark energy equation-of-state parameter $w$. For IDE models, we further sample the interaction parameter $\xi$. The posterior distributions are explored with \texttt{emcee}~\cite{Mackey}.

We compare these models using the Akaike Information Criterion (AIC) and the Bayesian Information Criterion (BIC), defined as
\begin{equation}
\mathrm{AIC} = -2\ln \mathcal{L}_{\max} + 2p, 
\qquad
\mathrm{BIC} = -2\ln \mathcal{L}_{\max} + p\ln N ,
\end{equation}
where $\mathcal{L}_{\max}$ is the maximum likelihood, $p$ and $N$ denote the number of free parameters and data points, respectively. In this work, $\mathcal{L}_{\max}$ denotes the maximum data likelihood from the GRB and SNe Ia likelihoods. The Gaussian prior on $\omega_{\rm b}$ is used for posterior sampling, but is not included in the AIC/BIC values. The number of free parameters $p$ includes all sampled cosmological and nuisance parameters, including $\omega_{\rm b}$ and $\sigma_{\rm res,\mu}$. We report the differences $\Delta\mathrm{AIC}$ and $\Delta\mathrm{BIC}$ relative to $\Lambda$CDM.

\section{Results}
\subsection{Full GRB covariance}
\begin{figure}[htbp]
\centering
\includegraphics[width=0.90\textwidth]{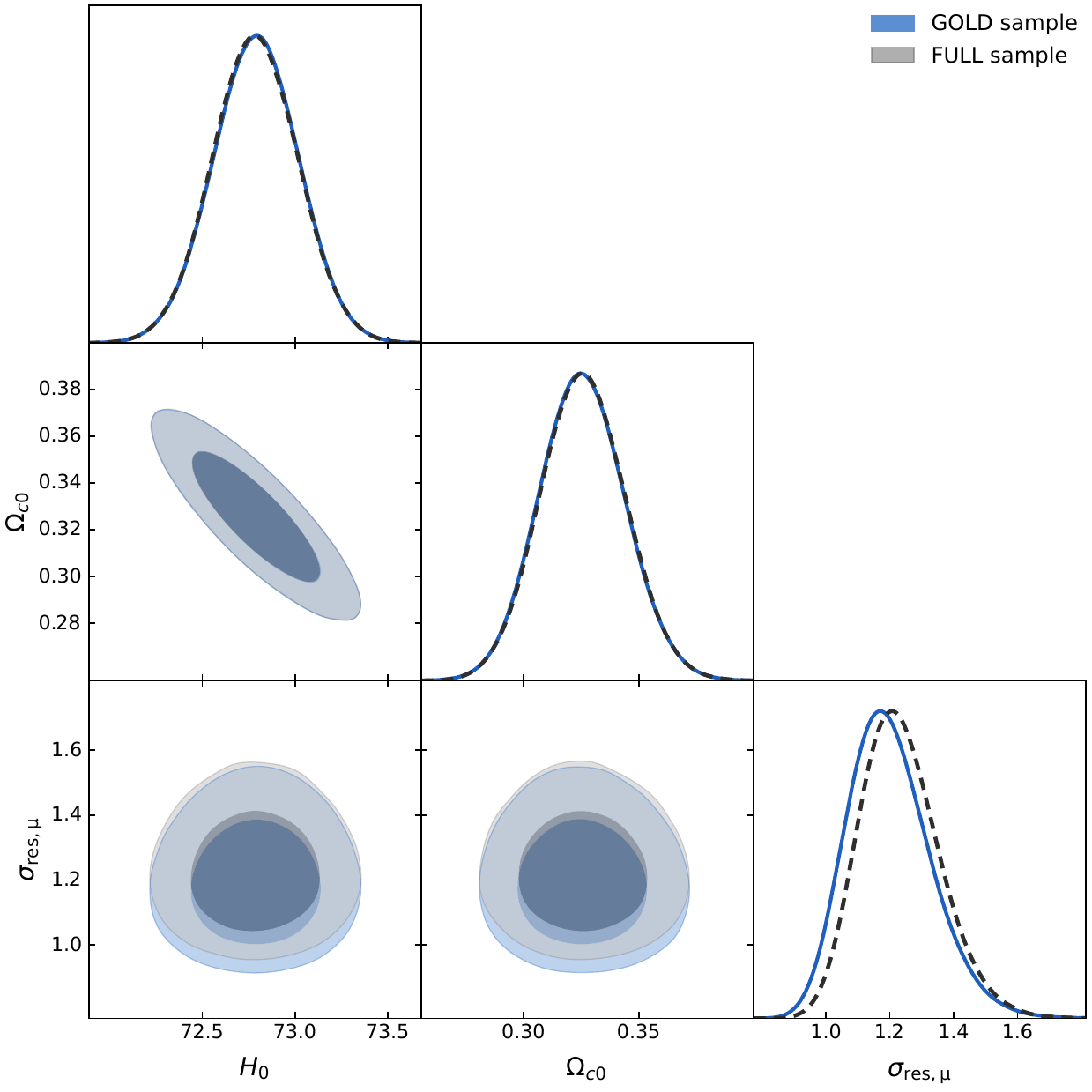}
\caption{Full-covariance constraints on cosmological parameters (68\% and 95\% credible regions) from the GOLD (blue) and FULL (gray) samples for the $\Lambda$CDM model.}
\label{fig:lcdm}
\end{figure}

\begin{figure}[htbp]
\centering
\includegraphics[width=0.90\textwidth]{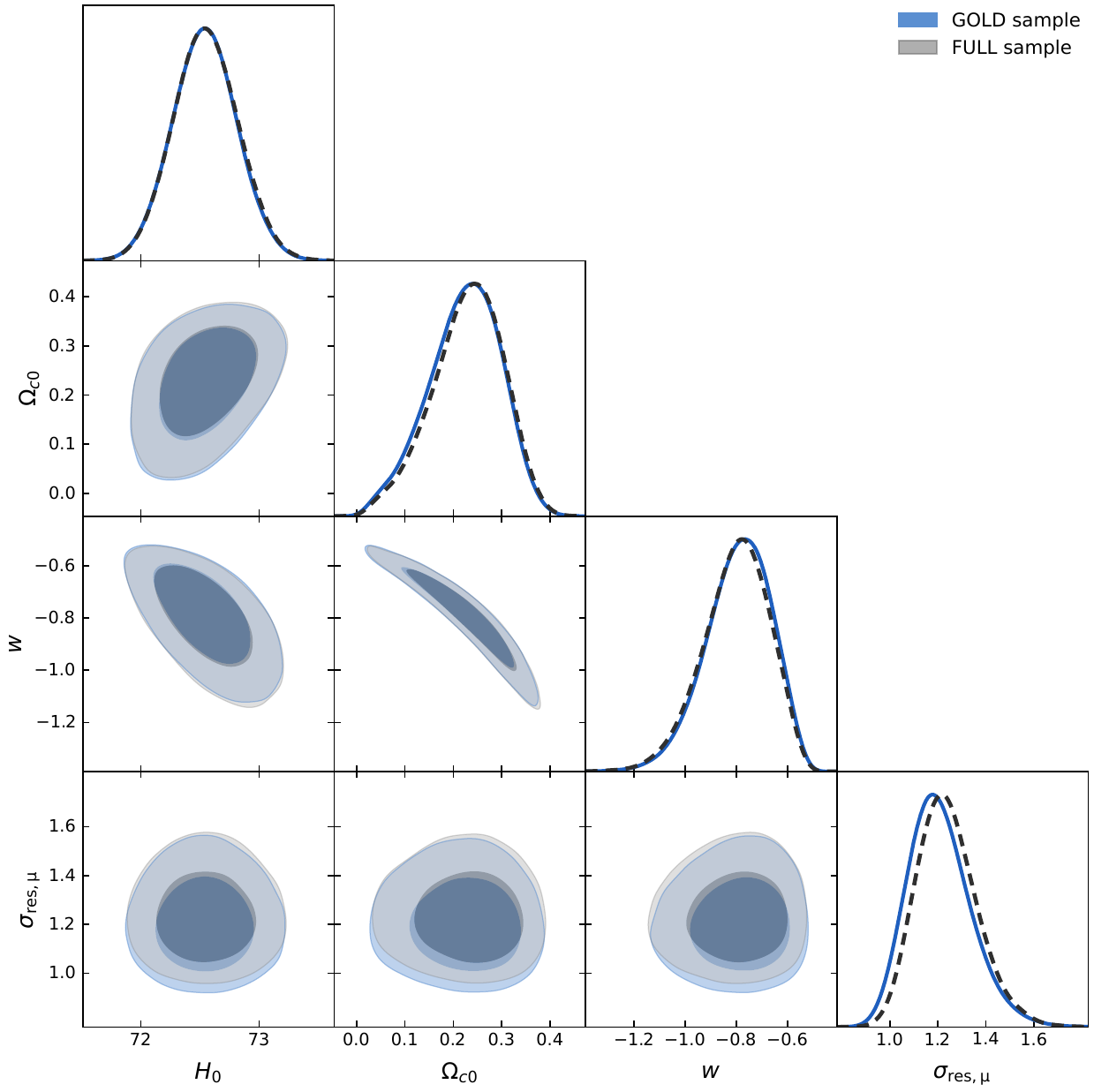}
\caption{Full-covariance constraints on cosmological parameters (68\% and 95\% credible regions) from the GOLD (blue) and FULL (gray) samples for the $w$CDM model.}
\label{fig:wcdm}
\end{figure}

\begin{figure}[htbp]
\centering
\includegraphics[width=0.90\textwidth]{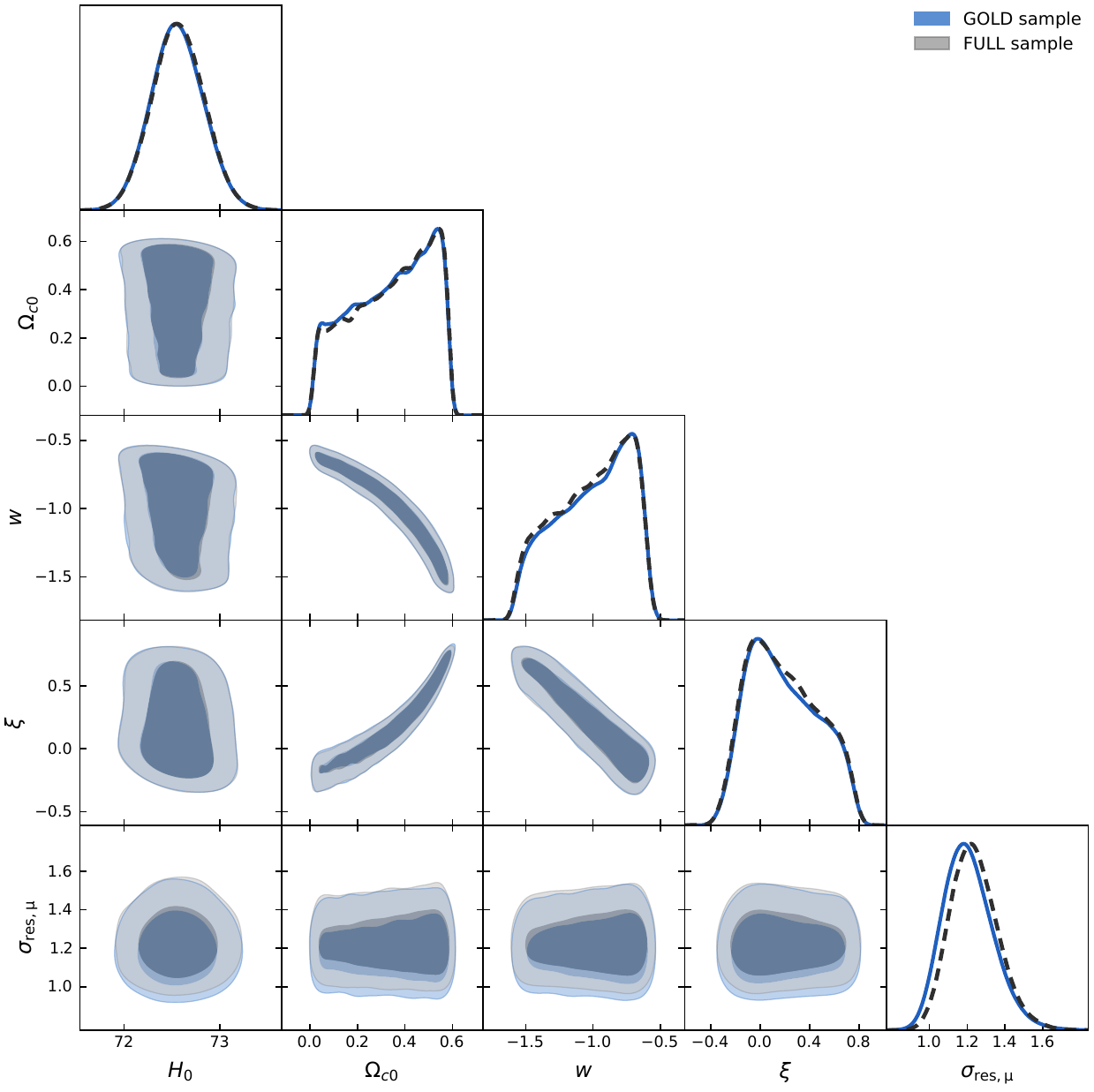}
\caption{Full-covariance constraints on cosmological parameters (68\% and 95\% credible regions) from the GOLD (blue) and FULL (gray) samples for the IDE-$\rho_{\rm de}$ model.}
\label{fig:ide_rhode}
\end{figure}

\begin{figure}[htbp]
\centering
\includegraphics[width=0.90\textwidth]{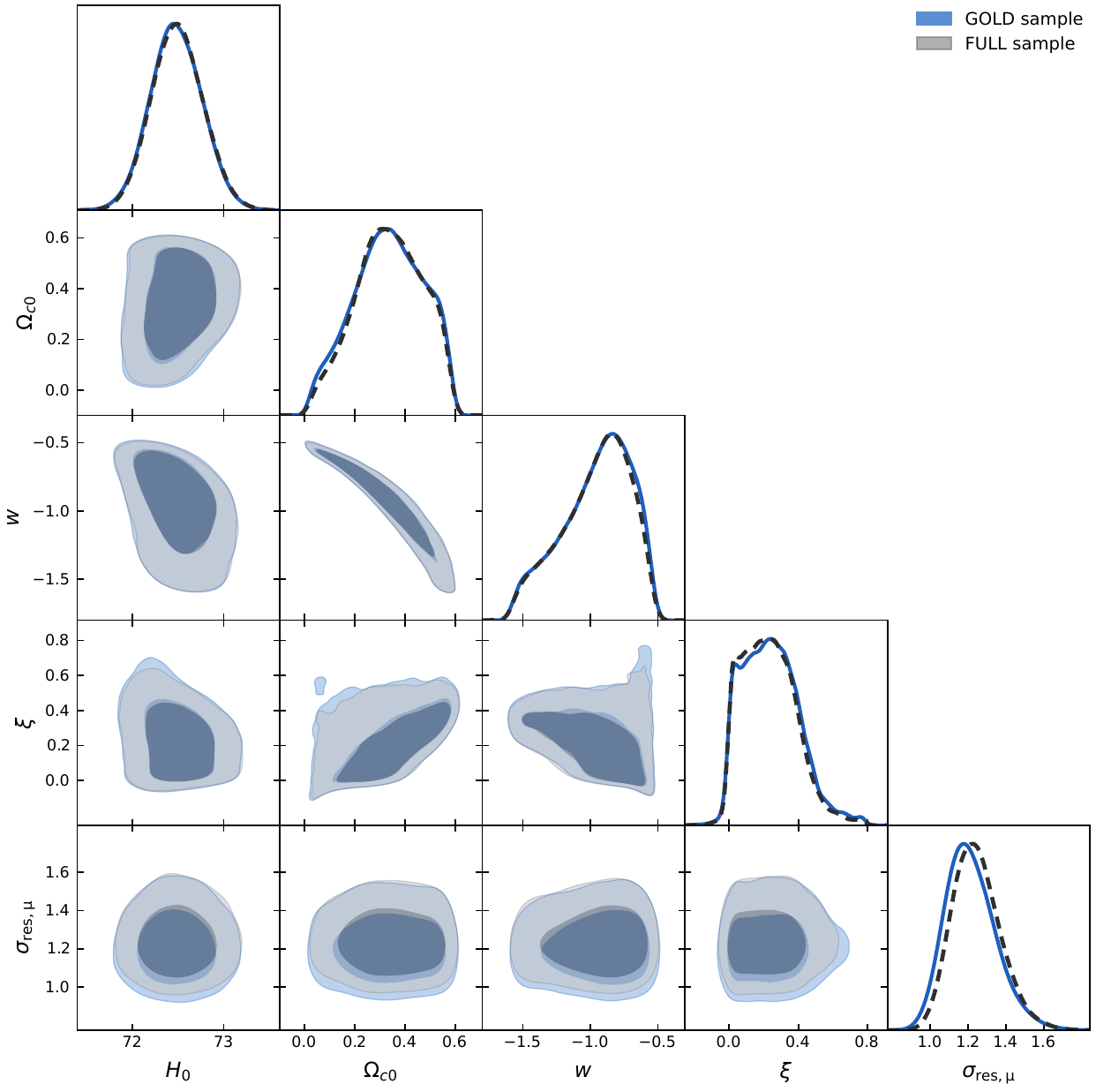}
\caption{Full-covariance constraints on cosmological parameters (68\% and 95\% credible regions) from the GOLD (blue) and FULL (gray) samples for the IDE-$\rho_{\rm c}$ model.}
\label{fig:ide_rhoc}
\end{figure}

The marginalized posterior medians and 68\% credible intervals for the cosmological parameters using the full GRB covariance are summarized in Table~\ref{tab:full_cov}. The 1D marginalized posterior distributions and 2D contours for the $\Lambda$CDM, $w$CDM, IDE-$\rho_{\rm de}$, and IDE-$\rho_{\rm c}$ models are shown in Figures~\ref{fig:lcdm}--\ref{fig:ide_rhoc}, respectively.

\begin{table}[htbp]
\centering
\caption{Marginalized posterior medians and 68\% credible intervals for the full GRB covariance. The $\Delta \mathrm{AIC}$ and $\Delta \mathrm{BIC}$ are relative to $\Lambda$CDM.}
\label{tab:full_cov}
\scriptsize
\setlength{\tabcolsep}{3.5pt}
\renewcommand{\arraystretch}{1.20}
\begin{adjustbox}{max width=\textwidth}
\begin{tabular}{@{}llccccccc@{}}
\toprule
Dataset & Models 
& $H_0$ 
& $\Omega_{\rm c0}$ 
& $w$ 
& $\xi$ 
& $\sigma_{\rm res,\mu}$ 
& $\Delta \mathrm{AIC}$ 
& $\Delta \mathrm{BIC}$ \\
\midrule

\multirow{4}{*}{GOLD sample}
& $\Lambda$CDM 
& $72.789^{+0.227}_{-0.229}$ 
& $0.325^{+0.019}_{-0.018}$ 
& -- 
& -- 
& $1.190^{+0.139}_{-0.118}$ 
& -- 
& -- \\

& $w$CDM 
& $72.542^{+0.272}_{-0.269}$ 
& $0.228^{+0.070}_{-0.082}$ 
& $-0.784^{+0.120}_{-0.137}$ 
& -- 
& $1.198^{+0.140}_{-0.120}$ 
& $-0.412$ 
& $5.061$ \\

& IDE-$\rho_{\rm de}$ 
& $72.554^{+0.274}_{-0.268}$ 
& $0.359^{+0.163}_{-0.218}$ 
& $-0.966^{+0.260}_{-0.360}$ 
& $0.161^{+0.361}_{-0.260}$ 
& $1.197^{+0.141}_{-0.120}$ 
& $1.588$ 
& $12.535$ \\

& IDE-$\rho_{\rm c}$ 
& $72.479^{+0.286}_{-0.279}$ 
& $0.336^{+0.150}_{-0.147}$ 
& $-0.915^{+0.222}_{-0.323}$ 
& $0.231^{+0.167}_{-0.164}$ 
& $1.205^{+0.143}_{-0.121}$ 
& $1.555$ 
& $12.503$ \\

\midrule

\multirow{4}{*}{FULL sample}
& $\Lambda$CDM 
& $72.784^{+0.229}_{-0.229}$ 
& $0.326^{+0.019}_{-0.018}$ 
& -- 
& -- 
& $1.224^{+0.132}_{-0.114}$ 
& -- 
& -- \\

& $w$CDM 
& $72.547^{+0.277}_{-0.270}$ 
& $0.234^{+0.068}_{-0.080}$ 
& $-0.793^{+0.121}_{-0.139}$ 
& -- 
& $1.230^{+0.133}_{-0.115}$ 
& $-0.220$ 
& $5.263$ \\

& IDE-$\rho_{\rm de}$ 
& $72.560^{+0.273}_{-0.266}$ 
& $0.363^{+0.161}_{-0.220}$ 
& $-0.975^{+0.264}_{-0.363}$ 
& $0.163^{+0.360}_{-0.267}$ 
& $1.229^{+0.132}_{-0.115}$ 
& $1.780$ 
& $12.747$ \\

& IDE-$\rho_{\rm c}$ 
& $72.487^{+0.286}_{-0.277}$ 
& $0.339^{+0.145}_{-0.140}$ 
& $-0.927^{+0.221}_{-0.314}$ 
& $0.218^{+0.163}_{-0.155}$ 
& $1.235^{+0.136}_{-0.117}$ 
& $1.750$ 
& $12.718$ \\

\bottomrule
\end{tabular}
\end{adjustbox}
\end{table}

\FloatBarrier

For the GOLD sample, the inferred Hubble constant remains close to $72.5$--$72.8~{\rm km\,s^{-1}\,Mpc^{-1}}$ in all models. Although the AIC value of the $w$CDM model is slightly lower than that of $\Lambda$CDM, the difference is too small to indicate a meaningful model preference. The BIC penalizes the additional parameters more strongly and clearly favors $\Lambda$CDM. The two IDE models have $\Delta{\rm BIC}$ values of about $12.5$, indicating that the GOLD sample does not support IDE. The FULL sample gives the same conclusion. The AIC difference between $w$CDM and $\Lambda$CDM remains small, while the BIC still favors $\Lambda$CDM. The IDE models remain penalized by the additional parameters.

The fitted effective residual scatter remains large when the full GRB covariance is used. For the $\Lambda$CDM model, we obtain $\sigma_{\rm res,\mu}\approx 1.20$ for the GOLD sample and $\sigma_{\rm res,\mu}\approx 1.23$ for the FULL sample. This indicates that, even after propagating the Amati-relation calibration covariance, the current GRB Hubble diagram still has substantial scatter. This scatter limits the ability of current GRBs to provide precise tests of dark-sector physics. Their main value lies in the high-redshift lever arm and in their ability to probe the shape of the expansion history.

\subsection{Diagonal GRB covariance}

The marginalized posterior medians and 68\% credible intervals for the diagonal GRB covariance are summarized in Table~\ref{tab:diag_cov}. The $\Delta \mathrm{AIC}$ and $\Delta \mathrm{BIC}$ are relative to $\Lambda$CDM. The marginalized posteriors for the diagonal GRB covariance are presented in \ref{app:diag_cov}, while supplementary numerical results not listed here are summarized in \ref{app:numerical_results}.

\begin{table}[htbp]
\centering
\caption{Marginalized posterior medians and 68\% credible intervals for the diagonal GRB covariance. The $\Delta \mathrm{AIC}$ and $\Delta \mathrm{BIC}$ are relative to $\Lambda$CDM.}
\label{tab:diag_cov}
\small
\setlength{\tabcolsep}{6pt}
\renewcommand{\arraystretch}{1.15}
\begin{adjustbox}{max width=\textwidth}
\begin{tabular}{@{}llcccc@{}}
\toprule
Dataset & Model & $H_0$ & $\sigma_{\rm res,\mu}$ & $\Delta \mathrm{AIC}$ & $\Delta \mathrm{BIC}$ \\
\midrule

\multirow{4}{*}{\begin{tabular}[c]{@{}c@{}}GOLD\\sample\end{tabular}}
& $\Lambda$CDM 
& $72.763^{+0.227}_{-0.229}$ 
& $1.420^{+0.168}_{-0.146}$ 
& -- 
& -- \\

& $w$CDM 
& $72.558^{+0.279}_{-0.272}$ 
& $1.449^{+0.173}_{-0.150}$ 
& $0.441$ 
& $5.915$ \\

& IDE-$\rho_{\rm de}$ 
& $72.576^{+0.273}_{-0.271}$ 
& $1.446^{+0.173}_{-0.151}$ 
& $2.441$ 
& $13.389$ \\

& IDE-$\rho_{\rm c}$ 
& $72.506^{+0.285}_{-0.283}$ 
& $1.477^{+0.178}_{-0.154}$ 
& $2.406$ 
& $13.354$ \\

\midrule

\multirow{4}{*}{\begin{tabular}[c]{@{}c@{}}FULL\\sample\end{tabular}}
& $\Lambda$CDM 
& $72.782^{+0.228}_{-0.230}$ 
& $1.600^{+0.173}_{-0.153}$ 
& -- 
& -- \\

& $w$CDM 
& $72.545^{+0.276}_{-0.270}$ 
& $1.627^{+0.175}_{-0.155}$ 
& $-0.143$ 
& $5.341$ \\

& IDE-$\rho_{\rm de}$ 
& $72.562^{+0.275}_{-0.269}$ 
& $1.627^{+0.176}_{-0.158}$ 
& $1.857$ 
& $12.825$ \\

& IDE-$\rho_{\rm c}$ 
& $72.490^{+0.286}_{-0.281}$ 
& $1.652^{+0.178}_{-0.158}$ 
& $1.827$ 
& $12.795$ \\

\bottomrule
\end{tabular}
\end{adjustbox}
\end{table}

\FloatBarrier

The diagonal approximation leads to a larger $\sigma_{\rm res,\mu}$, especially for the FULL sample, but it does not make IDE the preferred model. The BIC still favors the simpler $\Lambda$CDM model. The model ranking also remains stable for the FULL sample. This suggests that, when the off-diagonal calibration-induced correlations are ignored, the scatter parameter absorbs more of the residual structure in the GRB Hubble diagram.

The GOLD and FULL samples provide broadly consistent constraints on the main cosmological parameters. This is true for all four models, and it holds for both full and diagonal GRB covariance treatments. The agreement shows that the main cosmological conclusions are not driven by the choice of GOLD or FULL sample.

\subsection{Necessity of the effective residual scatter}

The fitted residual scatter $\sigma_{\rm res,\mu}$ is $1.19$--$1.24$ mag for the full covariance and $1.42$--$1.65$ mag for the diagonal covariance. The parameter $\sigma_{\rm res,\mu}$ is defined in distance-modulus space and is not directly equivalent to the intrinsic scatter of the Amati relation, which is defined in $\log_{10}E_{\rm iso}$ space and usually quoted in dex. Since $\mu_{\rm GRB}$ contains $2.5\log_{10}E_{\rm iso}$, the corresponding scale is approximately $0.48$--$0.50$ dex for the full covariance and $0.57$--$0.66$ dex for the diagonal covariance.

Table~\ref{tab:sigma0_fixed} gives the fits with $\sigma_{\rm res,\mu}=0$ for both samples, both covariance choices, and all four cosmologies. $\Delta\mathrm{AIC}_{\rm sc}$ and $\Delta\mathrm{BIC}_{\rm sc}$ compare each zero-scatter model with the corresponding free-scatter model. In every case, the zero-scatter likelihood provides a substantially worse fit than the corresponding free-scatter likelihood. Thus the propagated measurement and calibration covariance alone cannot adequately describe the GRB residuals.

When $\sigma_{\rm res,\mu}$ is fixed to zero in the full-covariance likelihood, the additional parameters of the extended cosmological models can partially reduce the $\chi^2_{\rm GRB}$. In particular, the zero-scatter IDE-$\rho_c$ model yields a lower $\chi^2_{\rm GRB}$ than the corresponding zero-scatter $\Lambda$CDM model, but its best-fit $\Omega_{c0}$ and $\xi$ reach their upper prior boundaries, while $\omega_b$ reaches its lower prior boundary. The best-fit $w$CDM model also reaches the lower prior boundary of $\Omega_{c0}$. This behavior suggests a compensatory fit, in which the cosmological parameters absorb residual structure that would otherwise be described by the effective residual-scatter term. The resulting model ranking should not be interpreted as evidence for IDE. The relevant comparison is that every zero-scatter model is decisively disfavored relative to its corresponding free-scatter model.

\begin{table}[htbp]
\centering
\caption{Fixed $\sigma_{\rm res,\mu}=0$ diagnostic. $\Delta\mathrm{AIC}_{\rm sc}$ and $\Delta\mathrm{BIC}_{\rm sc}$ compare each zero-scatter model with the corresponding
free-scatter model.}
\label{tab:sigma0_fixed}

\resizebox{\textwidth}{!}{%
\begin{tabular}{cclrrrrp{3.4cm}}
\toprule
Sample & Cov. & Model
& $-2\ln L$
& $\Delta\mathrm{AIC}_{\rm sc}$
& $\Delta\mathrm{BIC}_{\rm sc}$
& $\chi^2_{\rm GRB}/N$
& Boundary note \\
\midrule

\multirow[c]{8}{*}{GOLD}
& \multirow[c]{4}{*}{full}
& $\Lambda$CDM
& 8534.587 & 6681.5 & 6676.1 & 115.5 & -- \\

& & $w$CDM
& 8274.625 & 6424.0 & 6418.5 & 111.5
& $\Omega_{c0}$, $\omega_b$ low \\

& & IDE-$\rho_{\rm de}$
& 8258.310 & 6407.7 & 6402.2 & 111.2 & -- \\

& & IDE-$\rho_{\rm c}$
& 8197.882 & 6347.3 & 6341.8 & 110.3
& $\Omega_{c0}$, $\xi$ high, $\omega_b$ low \\

\cmidrule(lr){2-8}

& \multirow[c]{4}{*}{diag}
& $\Lambda$CDM
& 2225.643 & 355.7 & 350.2 & 8.3 & -- \\

& & $w$CDM
& 2212.449 & 344.1 & 338.6 & 7.9 & -- \\

& & IDE-$\rho_{\rm de}$
& 2212.449 & 344.1 & 338.6 & 7.9 & -- \\

& & IDE-$\rho_{\rm c}$
& 2212.448 & 344.1 & 338.6 & 7.9 & -- \\

\midrule

\multirow[c]{8}{*}{FULL}
& \multirow[c]{4}{*}{full}
& $\Lambda$CDM
& 7402.529 & 5504.2 & 5498.8 & 74.4 & -- \\

& & $w$CDM
& 7159.428 & 5263.4 & 5257.9 & 71.5
& $\Omega_{c0}$, $\omega_b$ low \\

& & IDE-$\rho_{\rm de}$
& 7143.533 & 5247.5 & 5242.0 & 71.3 & -- \\

& & IDE-$\rho_{\rm c}$
& 7087.894 & 5191.9 & 5186.4 & 70.6
& $\Omega_{c0}$, $\xi$ high, $\omega_b$ low \\

\cmidrule(lr){2-8}

& \multirow[c]{4}{*}{diag}
& $\Lambda$CDM
& 2352.786 & 421.7 & 416.2 & 7.9 & -- \\

& & $w$CDM
& 2339.862 & 410.9 & 405.4 & 7.6 & -- \\

& & IDE-$\rho_{\rm de}$
& 2339.862 & 410.9 & 405.4 & 7.6 & -- \\

& & IDE-$\rho_{\rm c}$
& 2339.860 & 410.9 & 405.5 & 7.6 & -- \\

\bottomrule
\end{tabular}%
}
\end{table}

\subsection{Robustness and interpretation of the constraints}

To identify what drives the joint constraints and assess their robustness, we examine the additional constraining power of the GRB data, the sensitivity of the inferred parameters to the Planck baryon-density prior, and the structure of the IDE posteriors.

\paragraph{Additional constraining power of the GRB data}
We use the SNe-only analysis as the reference for assessing the additional constraining power provided by the GRBs. Relative to $\Lambda$CDM, the SNe-only analysis gives $\Delta\mathrm{BIC}=3.05$ for $w$CDM and $\Delta\mathrm{BIC}\simeq10.48$ for both IDE models. The qualitative preference for $\Lambda$CDM is already present in the SNe data.

\begin{figure}[htbp]
\centering
\includegraphics[width=0.90\textwidth]{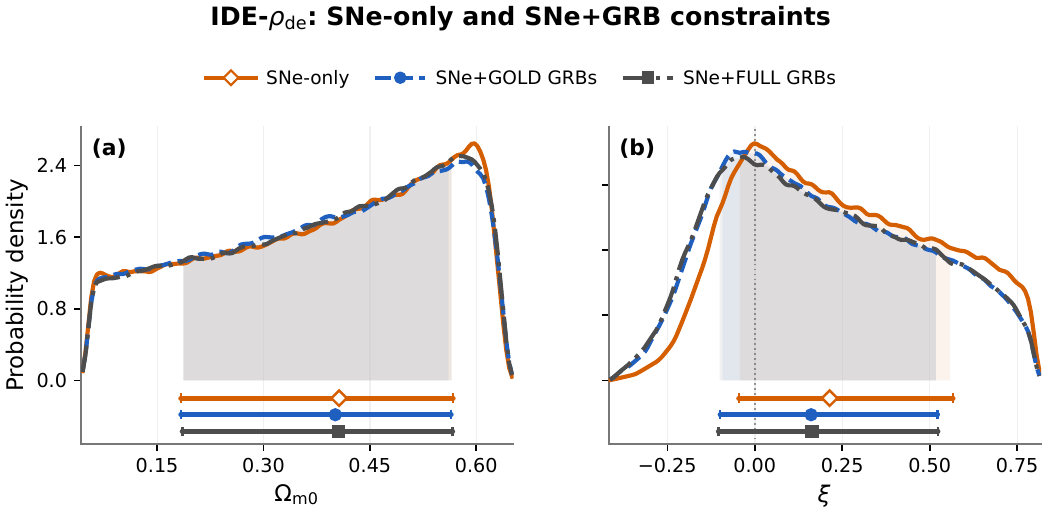}
\caption{Posterior probability densities of $\Omega_{\rm m0}$ and $\xi$ in the IDE-$\rho_{\rm de}$ model, obtained from the SNe-only analysis and the joint analyses combining Pantheon+ SNe with the GOLD or FULL GRB sample using the full GRB covariance. The points and horizontal bars show the medians and 68\% intervals, respectively, and the vertical dotted line marks the non-interacting limit $\xi=0$.}
\label{fig:fig5}
\end{figure}

\begin{figure}[htbp]
\centering
\includegraphics[width=0.90\textwidth]{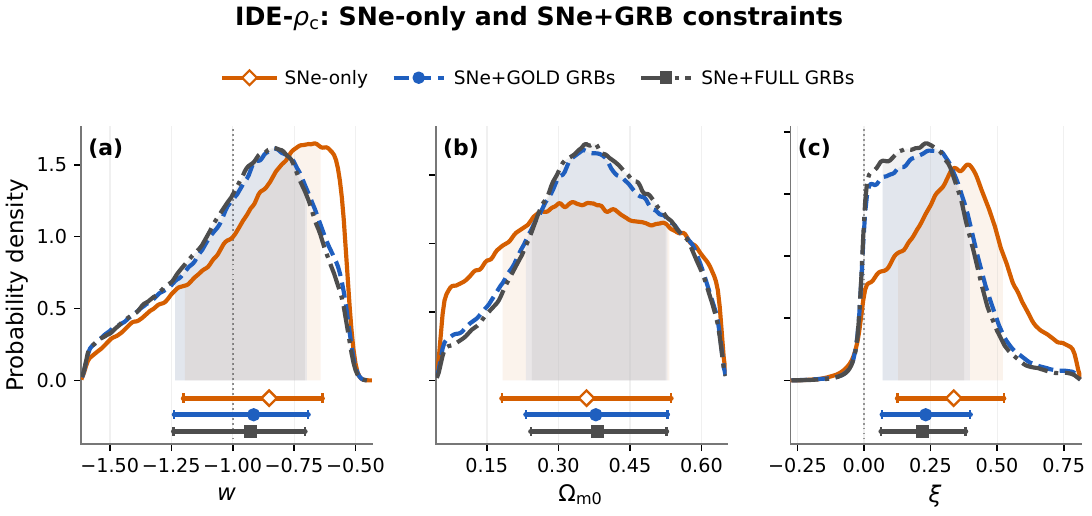}
\caption{Posterior probability densities of $w$, $\Omega_{\rm m0}$, and $\xi$ in the IDE-$\rho_{\rm c}$ model, obtained from the SNe-only analysis and the joint analyses combining Pantheon+ SNe with the GOLD or FULL GRB sample using the full GRB covariance. The points and horizontal bars show the medians and 68\% intervals, respectively, and the vertical dotted lines mark the reference values $w=-1$ and $\xi=0$.}
\label{fig:fig6}
\end{figure}

Table~\ref{tab:snonly}, Figure~\ref{fig:fig5} and Figure~\ref{fig:fig6} shows the percentage reduction in the 68\% interval width after adding the GRBs. The contribution is negligible for $\Lambda$CDM and IDE-$\rho_{\rm de}$. In $w$CDM, the $\Omega_m$ interval narrows by approximately 9--11\%, whereas the $w$ interval broadens by a similar fraction. The clearest improvement occurs for IDE-$\rho_{\rm c}$, for which the $\Omega_m$ and $\xi$ intervals narrow by approximately 16--20\%. The GRB data therefore tighten some parameter constraints but do not uniformly improve all constraints.

\begin{table}[htbp]
\centering
\caption{Percentage reduction in the 68\% interval width relative to SNe-only. Negative values indicate a broader joint interval.}
\label{tab:snonly}
\begin{tabular}{clrr}
\toprule
Model & Parameter & GOLD & FULL \\
\midrule
\multirow[c]{2}{*}{$\Lambda$CDM}
 & $H_0$      & 0.6  & 0.2  \\
 & $\Omega_m$ & 0.5  & 0.0  \\
\midrule
\multirow[c]{2}{*}{$w$CDM}
 & $w$        & -9.0 & -10.2 \\
 & $\Omega_m$ & 9.0  & 10.8  \\
\midrule
\multirow[c]{2}{*}{IDE-$\rho_{\rm de}$}
 & $\Omega_m$ & 0.7  & 0.8  \\
 & $\xi$      & -1.5 & -2.4 \\
\midrule
\multirow[c]{3}{*}{IDE-$\rho_{\rm c}$}
 & $w$        & 3.6  & 5.4  \\
 & $\Omega_m$ & 16.4 & 19.8 \\
 & $\xi$      & 17.1 & 20.4 \\
\bottomrule
\end{tabular}
\end{table}

\FloatBarrier

\paragraph{Sensitivity to the Planck baryon-density prior}

Removing the Planck prior leaves the inferred total matter density essentially unchanged. For $\Lambda$CDM and GOLD, the constraints are $\Omega_m=0.36745^{+0.01881}_{-0.01830}$ with the Planck prior and $\Omega_m=0.36740^{+0.01889}_{-0.01833}$ with the broad prior. The corresponding FULL results are $\Omega_m=0.36783^{+0.01881}_{-0.01847}$ and $\Omega_m=0.36779^{+0.01890}_{-0.01849}$, respectively. Across the four models, the median shifts in $H_0$ and $\sigma_{\rm res,\mu}$, as well as in $w$ and $\xi$ where applicable, are all below $0.03\sigma_{base}$, where $\sigma_{base}$ denotes the corresponding baseline 1$\sigma$ uncertainty. The conclusion that the data do not favor IDE is insensitive to the Planck baryon-density prior.

\paragraph{Structure of the IDE posteriors}
The IDE posteriors are broad and non-Gaussian, with pronounced degeneracies among $\Omega_{c0}$, $w$, and $\xi$. The no-interaction value, $\xi=0$, is not significantly excluded.

\section{Discussion}

Across all four cosmological models, the GOLD and FULL samples yield very similar constraints on the main cosmological parameters under both the full and diagonal GRB covariance treatments. In the fits that include the empirically required effective residual scatter, the IDE models do not improve likelihood sufficiently to compensate for the penalty associated with their additional parameters. This model-comparison conclusion is robust to the choice of GRB sample and covariance treatment and is also insensitive to the removal of the Planck prior on $\omega_b$. 

Several analyses have shown that the preference for IDE is sensitive to the assumed interaction term, the data combination, and the treatment of external calibration information~\cite{Valentino,Yang,Escamilla,Bensity,Giare,Litiannuo,Silva,Figueruelo}. Recent studies have further examined whether BAO data alter the allowed IDE parameter space or produce a mild preference for particular interaction forms~\cite{Giare,Litiannuo,Silva,Figueruelo}. GRB-based studies have also tested IDE and related dark-energy models using high-redshift GRB Hubble diagrams~\cite{Nong,Zhu}. Our results are consistent with the view that the GRB and SNe distance data considered here do not provide robust, model-independent evidence for a dark-sector interaction.

The SNe-only comparison clarifies the role of the GRBs. Most of the constraints on $H_0$ and $w$, as well as the baseline model ranking, are driven by SNe. The GRBs contribute a high-redshift lever arm and can moderately tighten some parameter constraints, especially those on $\Omega_m$ and $\xi$ in IDE-$\rho_{\rm c}$, but they do not dominate the joint constraints.

The zero-scatter diagnostic does not preserve the baseline model ranking under the full covariance and provides an unacceptable fit for every cosmology. The pronounced internal preference for flexible models, together with several best-fit parameters reaching the boundaries of their prior ranges, shows that model comparison depends on an adequate observational likelihood. Otherwise, a complex cosmological model can absorb residual structure that should be represented by the effective residual-scatter term.

There are still several limitations in the present analysis. The Amati relation is empirical, and the GRB effective residual-scatter remains large. The current likelihood includes the propagated covariance of the Amati intercept and slope, measurement errors, and one independent residual term, but not a population model for selection, detector thresholds, luminosity-function evolution, redshift incompleteness, redshift evolution of the Amati relation, host extinction, metallicity, or lensing~\cite{Butler,Ghirlan,LiLi,Schady,Perley}. Some of these effects may be absorbed by $\sigma_{\rm res,\mu}$, while correlated or redshift-dependent components cannot be represented by a single diagonal variance.

\section{Conclusions}

We have tested flat $\Lambda$CDM, $w$CDM, IDE-$\rho_{\rm de}$, and IDE-$\rho_c$ using high-redshift \textit{Fermi}/GBM GOLD and FULL GRBs together with Pantheon+ SNe Ia. The covariance of the low-redshift Amati calibration is propagated as a full non-diagonal matrix, and an effective residual scatter in GRB distance-modulus space is sampled jointly as a nuisance parameter.

With the residual scatter included, the full and diagonal covariance analyses give the same model-selection conclusion: neither IDE model improves the data likelihood enough to compensate for its additional parameters, and the BIC favors $\Lambda$CDM. Removing the Planck baryon-density prior leaves the IDE parameters and total matter density essentially unchanged. The SNe-only comparison shows that GRBs add limited precision for most parameters, with their clearest contribution being a roughly 17--20\% reduction of the $\xi$ interval in IDE-$\rho_c$.

Current GRB and Pantheon+ distance measurements provide no significant evidence for either of the two IDE interactions considered here. Future high-quality GRB observations, better low-redshift calibration samples, improved detector cross-calibration, and a more comprehensive treatment of GRB systematics will be required before GRBs can provide precision constraints competitive with those from SNe, BAO, or CMB measurements.

\section*{Acknowledgments}

This work was supported by the National SKA Program of China (Grants Nos. 2022SKA0110200 and 2022SKA0110203). This work was also supported by Xiaofeng Yang's ZHISHAN Distinguished Professor startup funding of Henan University.

\clearpage
\appendix

\section{Diagonal-covariance posterior plots}
\label{app:diag_cov}

\renewcommand{\thefigure}{A.\arabic{figure}}
\setcounter{figure}{0}

\begin{figure}[p]
\centering
\includegraphics[width=0.9\textwidth]{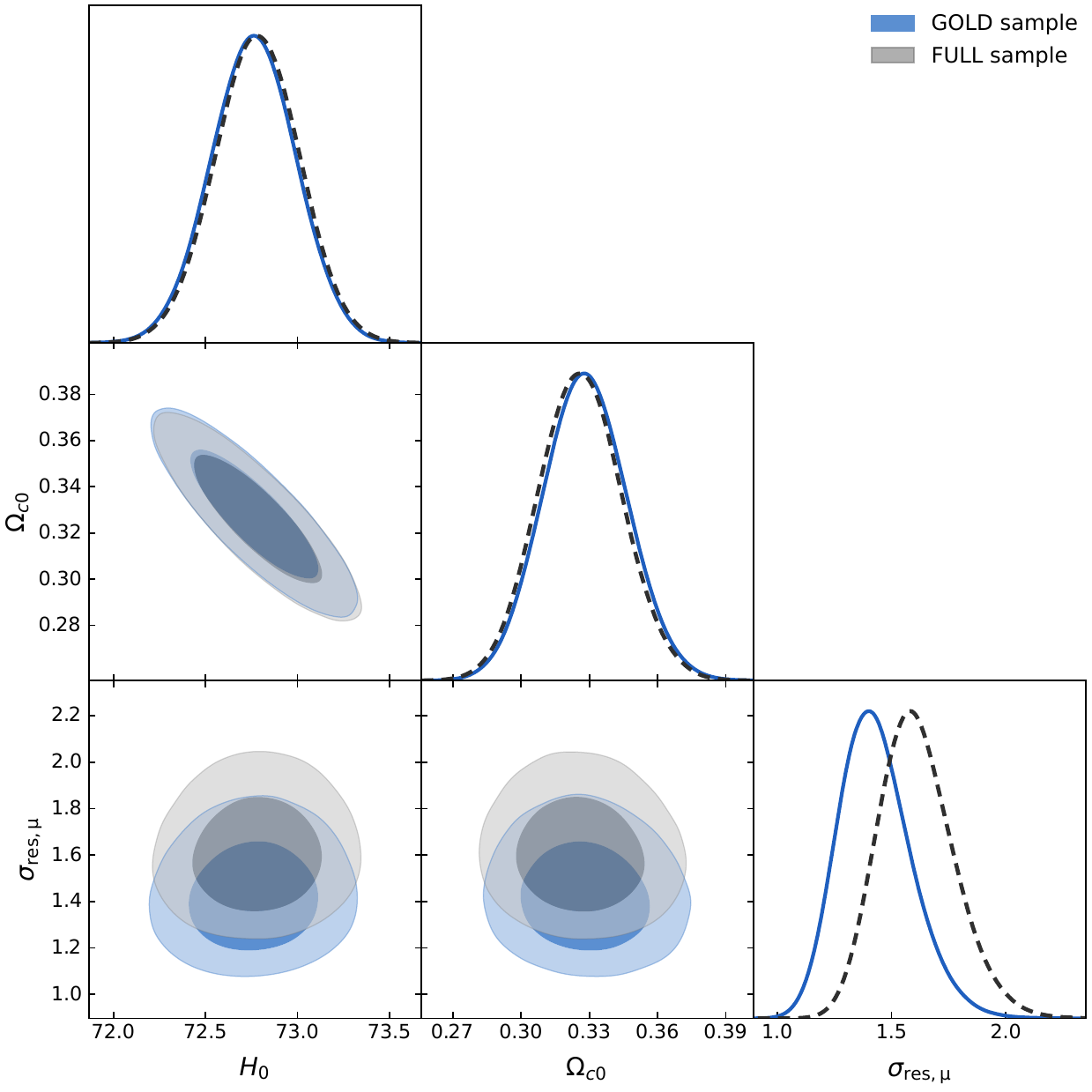}
\caption{Constraints on cosmological parameters (68\% and 95\% credible regions) for the $\Lambda$CDM model using the diagonal GRB covariance.}
\label{fig:diag_lcdm}
\end{figure}

\begin{figure}[p]
\centering
\includegraphics[width=0.9\textwidth]{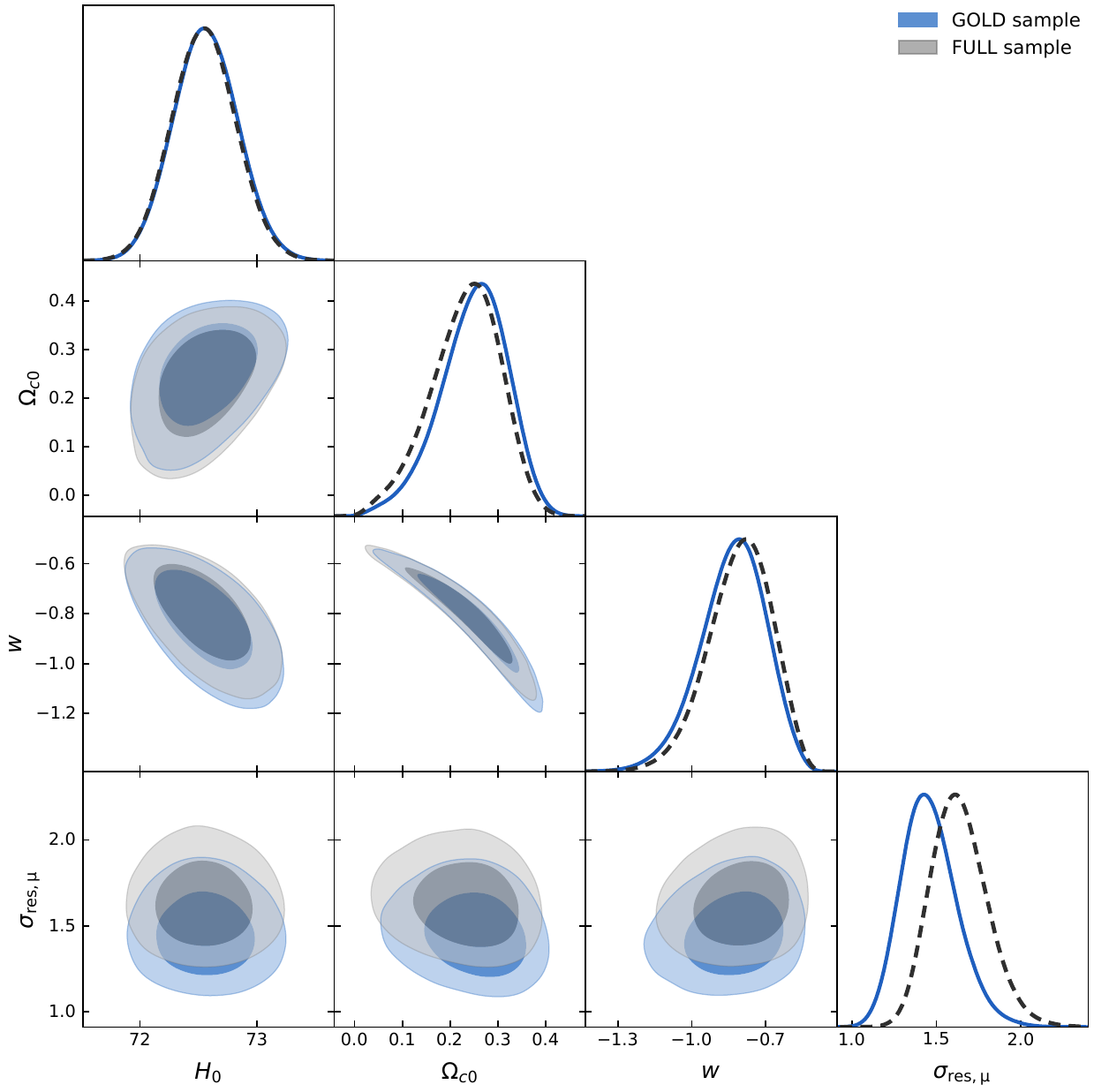}
\caption{Constraints on cosmological parameters (68\% and 95\% credible regions) for the $w$CDM model using the diagonal GRB covariance.}
\label{fig:diag_wcdm}
\end{figure}

\begin{figure}[p]
\centering
\includegraphics[width=0.9\textwidth]{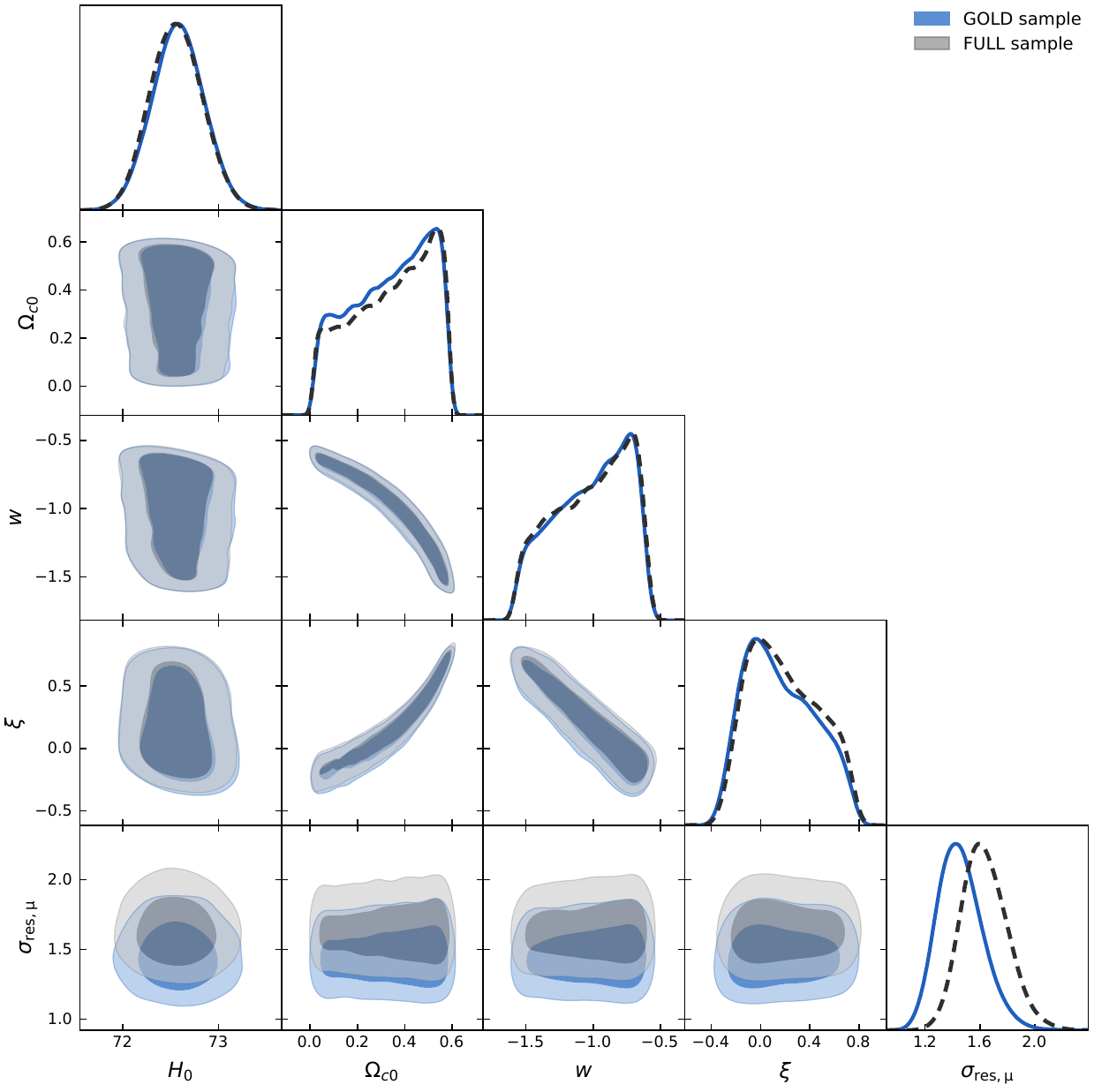}
\caption{Constraints on cosmological parameters (68\% and 95\% credible regions) for the IDE-$\rho_{\rm de}$ model using the diagonal GRB covariance.}
\label{fig:diag_ide_rhode}
\end{figure}

\begin{figure}[p]
\centering
\includegraphics[width=0.9\textwidth]{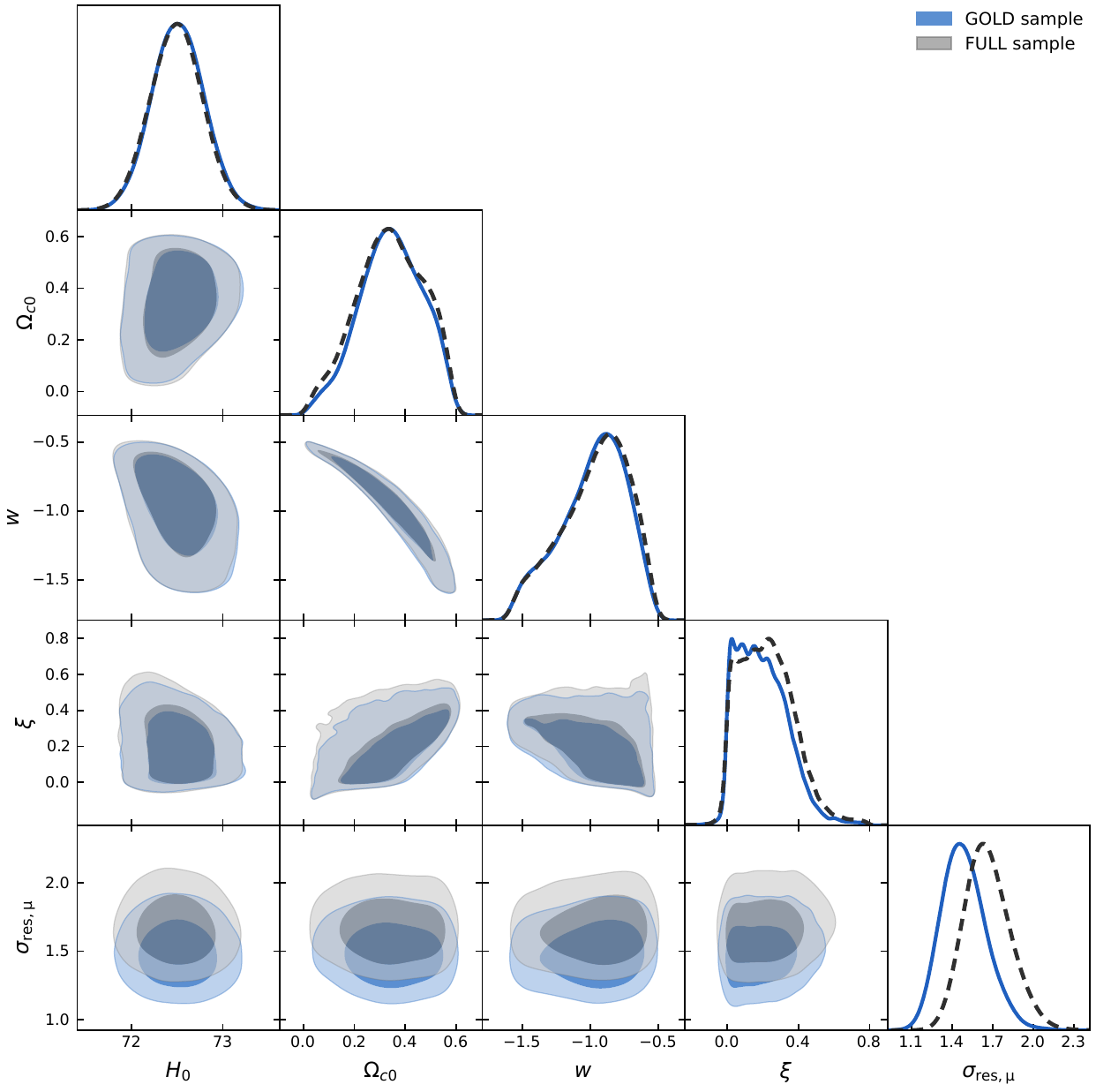}
\caption{Constraints on cosmological parameters (68\% and 95\% credible regions) for the IDE-$\rho_{\rm c}$ model using the diagonal GRB covariance.}
\label{fig:diag_ide_rhoc}
\end{figure}

The corresponding posterior plots for the diagonal GRB covariance are shown in Figures.~\ref{fig:diag_lcdm}--\ref{fig:diag_ide_rhoc}.

\clearpage

\section{Supplementary numerical results}
\label{app:numerical_results}

\renewcommand{\thetable}{B.\arabic{table}}
\setcounter{table}{0}

This appendix provides supplementary numerical results that are not fully reported in the baseline analysis.

\begin{table}[!htbp]
\centering
\caption{Supplementary numerical results for the full GRB covariance.}
\label{tab:full_cov_supplementary}
\begin{tabular}{llccc}
\toprule
Dataset & Models & $-2\ln \mathcal{L}_{\max}$ & AIC & BIC \\
\midrule
\multirow{4}{*}{GOLD sample}
& $\Lambda$CDM          & $1851.038$ & $1859.038$ & $1880.933$ \\
& $w$CDM                & $1848.626$ & $1858.626$ & $1885.994$ \\
& IDE-$\rho_{\rm de}$   & $1848.626$ & $1860.626$ & $1893.468$ \\
& IDE-$\rho_{\rm c}$    & $1848.593$ & $1860.593$ & $1893.435$ \\
\midrule
\multirow{4}{*}{FULL sample}
& $\Lambda$CDM          & $1896.291$ & $1904.291$ & $1926.226$ \\
& $w$CDM                & $1894.070$ & $1904.070$ & $1931.489$ \\
& IDE-$\rho_{\rm de}$   & $1894.070$ & $1906.070$ & $1938.973$ \\
& IDE-$\rho_{\rm c}$    & $1894.041$ & $1906.041$ & $1938.943$ \\
\bottomrule
\end{tabular}
\end{table}

\begin{table*}[!htbp]
\centering
\caption{Supplementary numerical results for the diagonal GRB covariance.}
\label{tab:diag_cov_supplementary}

\setlength{\tabcolsep}{3pt}
\renewcommand{\arraystretch}{1.15}
\scriptsize

\resizebox{\textwidth}{!}{%
\begin{tabular}{@{}llcccccc@{}}
\toprule
Dataset & Model & $\Omega_{\rm c0}$ & $w$ & $\xi$ 
& $-2\ln \mathcal{L}_{\max}$ & AIC & BIC \\
\midrule
\multirow{4}{*}{GOLD sample}
& $\Lambda$CDM        
& $0.328^{+0.019}_{-0.018}$ 
& --  
& --  
& $1867.924$ 
& $1875.924$ 
& $1897.819$ \\

& $w$CDM              
& $0.254^{+0.065}_{-0.076}$ 
& $-0.825^{+0.124}_{-0.142}$ 
& --  
& $1866.365$ 
& $1876.365$ 
& $1903.734$ \\

& IDE-$\rho_{\rm de}$ 
& $0.358^{+0.161}_{-0.216}$ 
& $-0.979^{+0.261}_{-0.357}$ 
& $0.131^{+0.361}_{-0.262}$ 
& $1866.365$ 
& $1878.365$ 
& $1911.207$ \\

& IDE-$\rho_{\rm c}$  
& $0.345^{+0.133}_{-0.126}$ 
& $-0.947^{+0.213}_{-0.298}$ 
& $0.183^{+0.159}_{-0.134}$ 
& $1866.330$ 
& $1878.330$ 
& $1911.172$ \\

\midrule
\multirow{4}{*}{FULL sample}
& $\Lambda$CDM        
& $0.326^{+0.018}_{-0.018}$ 
& --  
& --  
& $1929.094$ 
& $1937.094$ 
& $1959.029$ \\

& $w$CDM              
& $0.237^{+0.067}_{-0.080}$ 
& $-0.797^{+0.122}_{-0.138}$ 
& --  
& $1926.951$ 
& $1936.951$ 
& $1964.370$ \\

& IDE-$\rho_{\rm de}$ 
& $0.365^{+0.160}_{-0.223}$ 
& $-0.981^{+0.269}_{-0.364}$ 
& $0.164^{+0.363}_{-0.271}$ 
& $1926.951$ 
& $1938.951$ 
& $1971.854$ \\

& IDE-$\rho_{\rm c}$  
& $0.342^{+0.142}_{-0.137}$ 
& $-0.932^{+0.220}_{-0.310}$ 
& $0.213^{+0.161}_{-0.152}$ 
& $1926.922$ 
& $1938.922$ 
& $1971.824$ \\
\bottomrule
\end{tabular}%
}
\end{table*}

\clearpage

\bibliographystyle{elsarticle-num}
\bibliography{references}

\end{document}